\documentclass[preprint,12pt]{article}

\usepackage{amsmath,amssymb,array,calc,rotating,epsfig,psfrag, amscd, datetime, comment, color, bm}

\usepackage{color}
\usepackage[
      colorlinks=true,
      urlcolor=blue,    
      filecolor=blue,     
      citecolor=red,
      pdfstartview=FitV,
       bookmarksopen=true    
      ]{hyperref}
\usepackage[left=2.5cm,top=2.5cm,right=2.5cm,nohead]{geometry}

\newcommand{\nc}{\newcommand}

\definecolor{cardinal}{rgb}{0.6,0,0}
\definecolor{darkgreen}{rgb}{0,0.5,0}
\definecolor{golden}{rgb}{0.92, 0.7, 0}
\definecolor{midnight}{rgb}{0, 0, 0.5}
\definecolor{darkblue}{rgb}{0.2, 0, 0.8}
\definecolor{myblue}{rgb}{0.17, 0.13, .71}

\definecolor{slategrey}{rgb}{.44,.50,.56}
\definecolor{lightslategrey}{rgb}{.47,.53,.6}
\definecolor{silver}{rgb}{.75,.75,.75}
\nc{\SlateGrey}{\color{slategrey}}
\nc{\LightSlateGrey}{\color{lightslategrey}}
\nc{\Silver}{\color{silver}}

\definecolor{thistle}{RGB}{216,191,216}
\nc{\Thistle}{\color{thistle}}
\definecolor{blueviolet}{RGB}{138,43,226}
\nc{\BlueViolet}{\color{thistle}}
\definecolor{mediumpurple}{RGB}{147,112,219}
\nc{\MediumPurple}{\color{mediumpurple}}

\nc{\ra}{\rightarrow} 
\nc{\lra}{\leftrightarrow} 
\nc{\Ra}{\Rightarrow} 
\nc{\LRa}{\Leftightarrow} 
\nc{\blp}{{\big (}}
\nc{\brp}{{\big )}}
\nc{\Blp}{{\Big (}}
\nc{\Brp}{{\Big )}}
\nc{\bglp}{{\bigg (}}
\nc{\bgrp}{{\bigg )}}
\nc{\Bglp}{{\Bigg (}}
\nc{\Bgrp}{{\Bigg )}}
\nc{\slb}{{\rm [}}
\nc{\srb}{{\rm ]}}
\nc{\bslb}{{\rm \big [}}
\nc{\bsrb}{{\rm \big ]}}
\nc{\Bslb}{{\rm \Big [}}
\nc{\Bsrb}{{\rm \Big ]}}

\def\al{\alpha}

\def\eps{\epsilon}
\nc{\veps}{\varepsilon}
\def\gam{\gamma}

\def\lam{\lambda}
\def\om{\omega}

\nc{\vphi}{\varphi}
\def\tha{\theta}

\def\sig{\sigma}

\def\Gam{\Gamma}

\def\Lam{\Lambda}
\def\Om{\Omega}
\def\Sig{\Sigma}

\def\coeff#1#2{\relax{\textstyle {#1 \over #2}}\displaystyle}

\nc{\myvspace}{\rule[-1em]{0pt}{2.5em}}
\nc{\bea}{\begin{eqnarray}}
\nc{\eea}{\end{eqnarray}}
\nc{\be}{\begin{equation}}
\nc{\ee}{\end{equation}}
\nc{\barr}{\begin{array}}
\nc{\earr}{\end{array}}

\nc{\co}{{\cal o}}

\nc{\cA}{{\cal A}}
\nc{\cB}{{ \cal B}}
\def\cC{{\cal C}}
\def\cD{{\cal D}}

\nc{\cF}{{\cal F}}
\nc{\cG}{{\cal G}}
\def\cH{{\cal H}}
\def\cI{{\cal I}}

\def\cK{{\cal K}}
\nc{\cL}{{\cal L}}
\nc{\cM}{{\cal M}}

\def\cN{{\cal N}}

\def\cO{{\cal O}}

\nc{\cQ}{{\cal Q}}
\nc{\cR}{{\cal R}}
\def\cS{{\cal S}}
\def\cT{{\cal T}}

\def\cV{{\cal V}}
\def\cV{{\cal V}}
\def\cW{{\cal W}}

\def\cZ{{\cal Z}}
\nc{\cQd}{\cQ^{\dagger}}
\nc{\cRd}{\cR^{\dagger}}
\nc{\BB}{{\mathbb B}}
\nc{\CC}{{\mathbb C}}
\nc{\DD}{{\mathbb D}}
\nc{\EE}{{\mathbb E}}
\nc{\FF}{{\mathbb F}}
\nc{\GG}{{\mathbb G}}
\nc{\HH}{{\mathbb H}}
\nc{\JJ}{{\mathbb J}}
\nc{\MM}{{\mathbb M}}
\nc{\NN}{{\mathbb N}}
\nc{\PP}{{\mathbb P}}
\nc{\QQ}{{\mathbb Q}}
\nc{\RR}{{\mathbb R}}
\nc{\UU}{{\mathbb U}}
\nc{\ZZ}{{\mathbb Z}}
\nc{\calone}{{\mathbb 1}}

\nc{\half}{\frac{1}{2}}
\nc{\quarter}{\coeff{1}{4}}
\nc{\del}{\partial}

\nc{\delbar}{\bar\partial}
\nc{\thalf}{\frac{t}{2}}
\nc{\Spin}{\operatorname{Spin}}
\nc{\SO}{\operatorname{SO}}

\nc{\Sp}{{\rm Sp}}
\nc{\com}[2]{{ \left[ #1, #2 \right] }}
\nc{\acom}[2]{{ \left\{ #1, #2 \right\} }}
\nc{\rr}{\rightarrow}
\nc{\p}{\partial}
\nc{\LT}{{\LL_\T}}
\nc{\Tr}{{\rm Tr}}
\nc{\tr}{{\rm tr}}
\nc{\Adag}{A^{\dagger}}
\nc{\AdagI}{A^{\dagger I}}
\nc{\AdagJ}{A^{\dagger J}}
\nc{\AdagK}{A^{\dagger K}}
\nc{\AdagL}{A^{\dagger L}}
\nc{\AdagM}{A^{\dagger M}}
\nc{\Bdag}{B^{\dagger}}
\nc{\BdagI}{B^{\dagger}_I}
\nc{\BdagJ}{B^{\dagger}_J}
\nc{\BdagK}{B^{\dagger}_K}
\nc{\BdagL}{B^{\dagger}_L}
\nc{\BdagM}{B^{\dagger}_M}
\nc{\Cdag}{C^{\dagger}}
\nc{\CdagI}{C^{\dagger I}}
\nc{\CdagJ}{C^{\dagger J}}
\nc{\CdagK}{C^{\dagger K}}
\nc{\Ddag}{D^{\dagger}}
\nc{\DdagI}{D^{\dagger I}}
\nc{\DdagJ}{D^{\dagger J}}
\nc{\DdagK}{D^{\dagger K}}
\nc{\bva}{\breve{a}}
\nc{\bvb}{\breve{b}}
\nc{\bvc}{\breve{c}}
\nc{\bvd}{\breve{d}}
\nc{\bve}{\breve{e}}
\nc{\bvf}{\breve{f}}
\nc{\bvg}{\breve{g}}
\nc{\bvh}{\breve{h}}
\nc{\bvi}{\breve{i}}
\nc{\bvj}{\breve{j}}
\nc{\bvk}{\breve{k}}
\nc{\bvl}{\breve{l}}
\nc{\bvm}{\breve{m}}
\nc{\bvn}{\breve{n}}
\nc{\bvo}{\breve{o}}
\nc{\bvp}{\breve{p}}
\nc{\brvq}{\breve{q}}
\nc{\bvr}{\breve{r}}
\nc{\bvs}{\breve{s}}
\nc{\bvt}{\breve{t}}
\nc{\bvu}{\breve{u}}
\nc{\bvv}{\breve{v}}
\nc{\bvw}{\breve{w}}
\nc{\bvx}{\breve{x}}
\nc{\bvy}{\breve{y}}
\nc{\bvz}{\breve{z}}

\nc{\bvA}{\breve{A}}
\nc{\bvB}{\breve{B}}
\nc{\bvC}{\breve{C}}
\nc{\bvD}{\breve{D}}
\nc{\bvE}{\breve{E}}
\nc{\bvF}{\breve{F}}
\nc{\bvG}{\breve{G}}
\nc{\bvH}{\breve{H}}
\nc{\bvI}{\breve{I}}
\nc{\bvJ}{\breve{J}}
\nc{\bvK}{\breve{K}}
\nc{\bvL}{\breve{L}}
\nc{\bvM}{\breve{M}}
\nc{\bvN}{\breve{N}}
\nc{\bvO}{\breve{O}}
\nc{\bvP}{\breve{P}}
\nc{\bvQ}{\breve{Q}}
\nc{\bvR}{\breve{R}}
\nc{\bvS}{\breve{S}}
\nc{\bvT}{\breve{T}}
\nc{\bvU}{\breve{U}}
\nc{\bvV}{\breve{V}}
\nc{\bvcV}{\breve{\cV}}
\nc{\bvW}{\breve{W}}
\nc{\bvX}{\breve{X}}
\nc{\bvY}{\breve{Y}}
\nc{\bvZ}{\breve{Z}}

\nc{\ul}[1]{{\underline{#1}}}

\nc{\tal}{\widetilde{\alpha}}
\nc{\tbeta}{\widetilde{\beta}}
\nc{\ttha}{\tilde{\theta}}
\nc{\ttau}{\tilde{\tau}}
\nc{\tTha}{\tilde{\Theta}}
\nc{\tphi}{\tilde{\phi}}
\nc{\tsig}{\tilde{\sig}}
\nc{\tom}{\widetilde{\om}}
\nc{\tOm}{\widetilde{\Om}}
\nc{\tlam}{\widetilde{\lam}}
\nc{\tLam}{\tilde{\Lam}}
\nc{\tSig}{\widetilde{\Sig}}
\nc{\tPhi}{\tilde{\Phi}}
\nc{\tPhibar}{\ol{\tPhi}}
\nc{\tPi}{\widetilde{\Pi}}
\nc{\tpsi}{\widetilde{\psi}}
\nc{\tPsi}{\tilde{\Psi}}
\nc{\tgam}{\widetilde{\gam}}
\nc{\tGam}{\widetilde{\Gam}}
\nc{\tzeta}{\tilde{\zeta}}
\nc{\tZeta}{\tilde{\Zeta}}
\nc{\teta}{\widetilde{\eta}}
\nc{\teps}{\tilde{\eps}}
\nc{\tveps}{\tilde{\veps}}
\nc{\tEta}{\tilde{\Eta}}
\nc{\tchi}{\tilde{\chi}}
\nc{\tChi}{\tilde{\Chi}}
\nc{\txi}{\tilde{\xi}}
\nc{\tXi}{\widetilde{\Xi}}
\nc{\tnu}{\tilde{\nu}}
\nc{\tmu}{\tilde{\mu}}

\nc{\ta}{\tilde a}
\nc{\tb}{\tilde b}
\nc{\tc}{\tilde c}
\nc{\te}{\tilde e}
\nc{\tf}{\widetilde f}
\nc{\tg}{\widetilde g}
\nc{\ti}{\tilde i}
\nc{\tj}{\tilde j}
\nc{\tk}{\widetilde k}
\nc{\tl}{\tilde l}
\nc{\tm}{\widetilde m}
\nc{\tn}{\tilde n}
\nc{\tp}{\tilde{p}}
\nc{\tq}{\widetilde{q}}
\nc{\trr}{{\tilde r}}
\nc{\ts}{{\tilde s}}
\nc{\tu}{{\tilde u}}
\nc{\tv}{{\tilde v}}
\nc{\tw}{{\tilde w}}
\nc{\tx}{{\tilde x}}
\nc{\ty}{{\tilde y}}
\nc{\tz}{\tilde z}
\nc{\tA}{{\widetilde A}}
\nc{\tAbar}{{\ol \tA}}
\nc{\tB}{{\widetilde B}}
\nc{\tC}{{\widetilde C}}
\nc{\tD}{{\widetilde D}}
\nc{\tE}{{\widetilde E}}
\nc{\tF}{{\widetilde F}}
\nc{\tG}{{\widetilde G}}
\nc{\tcG}{{\widetilde \cG}}
\nc{\tH}{{\widetilde H}}
\nc{\tcH}{{\widetilde \cH}}
\nc{\tI}{{\widetilde I}}
\nc{\tcI}{{\widetilde \cI}}
\nc{\tJ}{{\widetilde J}}
\nc{\tJbar}{{\ol {\tilde J}}}
\nc{\tK}{{\widetilde K}}
\nc{\tL}{{\widetilde L}}
\nc{\tcL}{{\widetilde \cL}}
\nc{\tcLbar}{{\ol \tcL}}
\nc{\tM}{{\widetilde M}}
\nc{\tN}{{\widetilde N}}
\nc{\tcN}{{\widetilde \cN}}
\nc{\tP}{{\widetilde P}}
\nc{\tQ}{{\widetilde Q}}
\nc{\tR}{{\widetilde R}}
\nc{\tcR}{{\widetilde \cR}}
\nc{\tS}{\widetilde{S}}
\nc{\tT}{\widetilde{T}}
\nc{\tU}{\widetilde{U}}
\nc{\tUU}{\widetilde{\UU}}
\nc{\tV}{\widetilde{V}}
\nc{\tcVbar}{\ol{\widetilde{\cV}}}
\nc{\tW}{\widetilde{W}}
\nc{\tcF}{\widetilde{{\cal F}}}
\nc{\tX}{\widetilde{X}}
\nc{\tY}{\widetilde{Y}}
\nc{\tcZ}{\tilde{\cZ}}
\nc{\tcZbar}{\ol{\tcZ}}

\nc{\ha}{\hat a}
\nc{\hb}{\hat b}
\nc{\hc}{\widehat c}
\nc{\hd}{\widehat d}
\nc{\he}{\widehat e}
\nc{\hf}{\widehat f}
\nc{\hg}{\widehat g}
\nc{\hh}{\widehat h}
\nc{\hn}{\widehat n}
\nc{\hp}{\widehat p}
\nc{\hq}{\widehat q}
\nc{\hr}{\widehat r}
\nc{\hs}{\widehat s}
\nc{\hu}{\widehat u}
\nc{\hv}{\widehat v}
\nc{\hw}{\widehat w}
\nc{\bhw}{{\bf \hw}}
\nc{\hx}{\widehat x}
\nc{\hy}{\widehat y}
\nc{\hz}{\widehat z}
\nc{\zhat}{\hat z}
\nc{\hA}{\widehat{A}}
\nc{\hB}{\widehat{B}}
\nc{\hC}{\widehat{C}}
\nc{\hD}{\widehat{D}}
\nc{\hcD}{\widehat{\cD}}
\nc{\hE}{\widehat{E}}
\nc{\hF}{\widehat{F}}
\nc{\hcF}{\widehat{\cF}}
\nc{\hG}{\widehat{G}}
\nc{\hcG}{\widehat{\cG}}
\nc{\hH}{\widehat{H}}
\nc{\hI}{\widehat{I}}
\nc{\hcI}{\widehat{\cI}}
\nc{\hJ}{\widehat{J}}
\nc{\hK}{\widehat{K}}
\nc{\hcK}{\widehat{\cK}}
\nc{\hL}{\widehat{L}}
\nc{\hcL}{\widehat{\cL}}
\nc{\hM}{\widehat M}
\nc{\hcM}{\widehat{\cM}}
\nc{\hN}{\widehat{N}}
\nc{\hO}{\widehat{O}}
\nc{\hcO}{\widehat{\cO}}
\nc{\hP}{\widehat{P}}
\nc{\hQ}{\widehat{Q}}
\nc{\hcQ}{\widehat{\cQ}}
\nc{\hcR}{\widehat{\cR}}
\nc{\hR}{\widehat{R}}
\nc{\hS}{\widehat{S}}
\nc{\hcS}{\widehat{\cS}}
\nc{\hT}{\widehat{T}}
\nc{\hU}{\widehat{U}}
\nc{\hV}{\widehat V}
\nc{\hcV}{\widehat \cV}
\nc{\tcV}{\widetilde{\cV}}
\nc{\hX}{\widehat X}
\nc{\hcZ}{\widehat \cZ}
\nc{\hcZbar}{\ol{\widehat \cZ}}

\nc{\heta}{\widehat{\eta}}
\nc{\hal}{\widehat \alpha}
\nc{\hbeta}{\widehat \beta}
\nc{\heps}{\widehat \eps}
\nc{\hphi}{\widehat{\phi}}
\nc{\hkap}{\hat{\kappa}}
\nc{\hchi}{\widehat{\chi}}
\nc{\hpsi}{\widehat{\psi}}
\nc{\hgam}{\widehat{\gam}}
\nc{\hPhi}{\hat{\Phi}}
\nc{\hPsi}{\hat{\Psi}}
\nc{\hGam}{\hat{\Gam}}
\nc{\omhat}{\widehat{\om}}
\nc{\Omhat}{\widehat{\Om}}
\nc{\hsig}{\widehat{\sig}}
\nc{\hSig}{\widehat{\Sig}}
\nc{\htha}{\hat{\tha}}
\nc{\hrho}{\widehat{\rho}}
\nc{\hdel}{\widehat{\del}}
\nc{\hdelbar}{\ol{\hdel}}
\nc{\hnabla}{\widehat{\nabla}}

\nc{\w}{\wedge}

\nc{\vb}{\vec b}
\nc{\vc}{\vec c}
\nc{\vd}{\vec d}
\nc{\ve}{\vec e}
\nc{\vf}{\vec f}
\nc{\vg}{\vec g}
\nc{\vh}{\vec h}
\nc{\vk}{\vec k}
\nc{\vl}{\vec l}
\nc{\vm}{\vec m}
\nc{\vn}{\vec n}
\nc{\vp}{\vec p}
\nc{\vq}{\vec q}
\nc{\vr}{\vec r}
\nc{\vs}{\vec s}
\nc{\vv}{\vec v}
\nc{\vw}{\vec w}
\nc{\vx}{\vec x}
\nc{\vy}{\vec y}
\nc{\vz}{\vec z}

\nc{\vA}{\vec A}
\nc{\vB}{\vec B}
\nc{\vC}{\vec C}
\nc{\vD}{\vec D}
\nc{\vE}{\vec E}
\nc{\vF}{\vec F}
\nc{\vG}{\vec G}
\nc{\vH}{\vec H}
\nc{\vP}{\vec P}
\nc{\vQ}{\vec Q}
\nc{\vR}{\vec R}
\nc{\vS}{\vec S}
\nc{\vV}{\vec V}
\nc{\vW}{\vec W}
\nc{\vX}{\vec X}
\nc{\vY}{\vec Y}
\nc{\vZ}{\vec Z}

\nc{\val}{\vec \al}
\nc{\vbeta}{\vec \beta}
\nc{\vmu}{\vec \mu}
\nc{\vtha}{\vec \theta}
\nc{\vecphi}{\vec \phi}
\nc{\vecvphi}{\vec \vphi}
\nc{\vecsig}{\vec \sig}
\nc{\vom}{\vec \om}

\nc{\ol}{\overline}
\nc{\abar}{\ol{a}}
\nc{\bbar}{\ol{b}}
\nc{\cbar}{\ol{c}}
\nc{\dbar}{\ol{d}}
\nc{\ebar}{\ol{e}}
\nc{\fbar}{\ol{f}}
\nc{\gbar}{\ol{g}}
\nc{\ibar}{\ol{\imath}}
\nc{\jbar}{\ol{\jmath}}
\nc{\kbar}{\ol{k}}
\nc{\lbar}{\ol{l}}
\nc{\mbar}{\ol{m}}
\nc{\nbar}{\ol{n}}
\nc{\pbar}{\ol{p}}
\nc{\qbar}{\ol{q}}
\nc{\rbar}{\ol{r}}
\nc{\sbar}{\ol{s}}
\nc{\ubar}{\ol{u}}
\nc{\vbar}{\ol{v}}
\nc{\wbar}{\ol{w}}
\nc{\xbar}{\ol{x}}
\nc{\ybar}{\ol{y}}
\nc{\zbar}{\ol{z}}

\nc{\Abar}{\ol{A}}
\nc{\cAbar}{\ol{\cA}}
\nc{\Bbar}{\ol{B}}
\nc{\cBbar}{\ol{\cB}}
\nc{\Cbar}{\ol{C}}
\nc{\cCbar}{\ol{\cC}}
\nc{\Dbar}{\ol{D}}
\nc{\Ebar}{\ol{E}}
\nc{\hEbar}{\ol{\hE}}
\nc{\Fbar}{\ol{F}}
\nc{\Gbar}{\ol{G}}
\nc{\Jbar}{\ol{J}}
\nc{\Kbar}{\ol{K}}
\nc{\cKbar}{\ol{\cK}}
\nc{\Lbar}{\ol{L}}
\nc{\cLbar}{\ol{\cL}}
\nc{\Mbar}{\ol{M}}
\nc{\Nbar}{\ol{N}}
\nc{\Pbar}{\ol{P}}
\nc{\Qbar}{\ol{Q}}
\nc{\Rbar}{\ol{R}}
\nc{\Sbar}{\ol{S}}
\nc{\Tbar}{\ol{T}}
\nc{\Ubar}{\ol{U}}
\nc{\Vbar}{\ol{V}}
\nc{\cVbar}{\ol{\cV}}
\nc{\Wbar}{\ol{W}}
\nc{\cWbar}{\ol{\cW}}
\nc{\Xbar}{{\overline X}}
\nc{\Ybar}{{\overline Y}}
\nc{\Zbar}{{\overline Z}}
\nc{\cZbar}{{\overline \cZ}}

\nc{\epsbar}{\ol{\epsilon}}
\nc{\albar}{\ol{\al}}
\nc{\Albar}{\ol{\Al}}
\nc{\betabar}{\ol{\beta}}
\nc{\Betabar}{\ol{\Beta}}
\nc{\deltabar}{\ol{\delta}}
\nc{\etabar}{\ol{\eta}}
\nc{\lambar}{\ol{\lambda}}
\nc{\kapbar}{\ol{\kappa}}
\nc{\zetabar}{\ol{\zeta}}
\nc{\Zetabar}{\ol{\Zeta}}
\nc{\taubar}{\ol{\tau}}
\nc{\Taubar}{\ol{\Tau}}
\nc{\psibar}{\ol{\psi}}
\nc{\Psibar}{\ol{\Psi}}
\nc{\tpsibar}{\ol{\tpsi}}
\nc{\tPsibar}{\ol{\tPsi}}
\nc{\phibar}{\ol{\phi}}
\nc{\Phibar}{\ol{\Phi}}
\nc{\chibar}{\ol{\chi}}
\nc{\sigbar}{\ol{\sig}}
\nc{\Sigbar}{\ol{\Sig}}
\nc{\mubar}{\ol{\mu}}
\nc{\nubar}{\ol{\nu}}
\nc{\rhobar}{\ol{\rho}}
\nc{\ombar}{\ol{\om}}
\nc{\Ombar}{\ol{\Om}}
\nc{\Deltabar}{\ol{\Delta}}
\nc{\Thetabar}{\ol{\Theta}}
\nc{\xibar}{\ol{\xi}}
\nc{\Xibar}{\ol{\Xi}}

\nc{\Dthbar}{\ol{\rm D3}}

\nc{\fdot}{\dot{f}}
\nc{\gdot}{\dot{g}}
\nc{\pdot}{\dot{p}}
\nc{\qdot}{\dot{q}}
\nc{\rdot}{\dot{r}}
\nc{\sdot}{\dot{s}}
\nc{\tdot}{\dot{t}}
\nc{\udot}{\dot{u}}
\nc{\vdot}{\dot{v}}
\nc{\wdot}{\dot{w}}
\nc{\xdot}{\dot{x}}
\nc{\xddot}{\ddot{x}}
\nc{\ydot}{\dot{y}}
\nc{\zdot}{\dot{z}}
\nc{\yddot}{\ddot{y}}

\nc{\Adot}{\dot{A}}
\nc{\Bdot}{\dot{B}}
\nc{\Cdot}{\dot{C}}
\nc{\dotD}{\dot{D}}
\nc{\Fdot}{\dot{F}}
\nc{\Pdot}{\dot{P}}
\nc{\Qdot}{\dot{Q}}
\nc{\Rdot}{\dot{R}}
\nc{\Sdot}{\dot{S}}
\nc{\Tdot}{\dot{T}}
\nc{\Udot}{\dot{U}}
\nc{\Vdot}{\dot{V}}
\nc{\Wdot}{\dot{W}}

\nc{\taudot}{\dot{\tau}}
\nc{\phidot}{\dot{\phi}}
\nc{\psidot}{\dot{\psi}}
\nc{\chidot}{\dot{\chi}}
\nc{\Gamdot}{\dot{\Gam}}
\nc{\sinp}{s_{\phi}}
\nc{\cosp}{c_{\phi}}
\nc{\tanp}{t_{\phi}}
\nc{\spone}{s_{\phi_1}}
\nc{\cpone}{c_{\phi_1}}
\nc{\tpone}{t_{\phi_1}}
\nc{\sptwo}{s_{\phi_2}}
\nc{\cptwo}{c_{\phi_2}}
\nc{\tptwo}{t_{\phi_2}}
\nc{\spth}{s_{\phi_3}}
\nc{\cpth}{c_{\phi_3}}
\nc{\tpth}{t_{\phi_3}}
\nc{\calp}{c_{\al}}
\nc{\salp}{s_{\al}}

\nc{\csch}{{\rm csch}}
\nc{\sech}{{\rm sech}}

\nc{\cothzlami}{\coth(z-\lam_i)}
\nc{\coshzlami}{\cosh(z-\lam_i)}
\nc{\sinhzlami}{\sinh(z-\lam_i)}

\nc{\cothzlamj}{\coth(z-\lam_j)}
\nc{\coshzlamj}{\cosh(z-\lam_j)}
\nc{\sinhzlamj}{\sinh(z-\lam_j)}

\nc{\cothlamij}{\coth(\lam_i-\lam_j)}
\nc{\coshlamij}{\cosh(\lam_i-\lam_j)}
\nc{\sinhlamij}{\sinh(\lam_i-\lam_j)}

\nc{\bah}{{\mathbf {\hat{A}}}}
\nc{\bX}{{\mathbf X}}
\nc{\ba}{{\bf a}}
\nc{\bb}{{\bf b}}
\nc{\bc}{{\bf c}}
\nc{\bd}{{\bf d}}
\nc{\bg}{{\bf g}}
\nc{\bk}{{\bf k}}
\nc{\bl}{{\bf l}}
\nc{\bn}{{\bf n}}
\nc{\bo}{{\bf o}}
\nc{\bp}{{\bf p}}
\nc{\bq}{{\bf q}}
\nc{\br}{{\bf r}}
\nc{\bs}{{\bf s}}
\nc{\bt}{{\bf t}}
\nc{\bu}{{\bf u}}
\nc{\bv}{{\bf v}}
\nc{\bw}{{\bf w}}
\nc{\bx}{{\bf x}}
\nc{\by}{{\bf y}}
\nc{\bz}{{\bf z}}

\nc{\bH}{{\bf H}}
\nc{\bP}{{\bf P}}
\nc{\bQ}{{\bf Q}}

\nc{\bom}{{\bf \om}}
\nc{\bombar}{{\mathbf \ombar}}
\nc{\bPhi}{{\bf \Phi}}
\nc{\bSig}{{\bf \Sig}}

\nc{\rma}{{\rm a}}
\nc{\rmb}{{\rm b}}
\nc{\rmc}{{\rm c}}
\nc{\rmd}{{\rm d}}
\nc{\rmg}{{\rm g}}
\nc{\rk}{{\rm k}}
\nc{\rml}{{\rm l}}
\nc{\rmm}{{\rm m}}
\nc{\rmn}{{\rm n}}
\nc{\rmo}{{\rm o}}
\nc{\rmp}{{\rm p}}
\nc{\rmq}{{\rm q}}
\nc{\rmr}{{\rm r}}
\nc{\rms}{{\rm s}}
\nc{\rmt}{{\rm t}}
\nc{\rmu}{{\rm u}}
\nc{\rmv}{{\rm v}}
\nc{\rmw}{{\rm w}}
\nc{\rmx}{{\rm x}}
\nc{\rmy}{{\rm y}}
\nc{\rmz}{{\rm z}}

\nc{\dal}{\dot{\al}}
\nc{\thadot}{\dot{\tha}}
\nc{\thab}{\bar{\theta}}
\nc{\thal}{\theta^{\al}}
\nc{\thdal}{\bar{\theta}^{\dal}}

\nc{\thsigthm}{\tha \sigma^m \thab}
\nc{\thsigthn}{\tha \sigma^n \thab}

\nc{\Dal}{D_{\al}}
\nc{\Ddal}{\bar{D}_{\dal}}
\nc{\CDal}{{\cal D}_{\al}}
\nc{\CDdal}{\bar{\cal D}_{\dal}}

\nc{\eq}[1]{{(\ref{#1})}}
\nc{\eqtwo}[2]{{(\ref{#1},\ref{#2})}}
\nc{\eqthree}[3]{(\ref{#1},\ref{#2},\ref{#3})}
\nc{\eqfour}[4]{(\ref{#1},\ref{#2},\ref{#3},\ref{#4})}
\nc{\eqfive}[5]{(\ref{#1},\ref{#2},\ref{#3},\ref{#4,\ref{#5}})}
\nc{\non}{\nonumber}
\nc{\Fzero}{F_{(0)}}
\nc{\Ftwo}{F_{(2)}}
\nc{\Ffour}{F_{(4)}}
\nc{\Fone}{F_{(1)}}
\nc{\Fthree}{F_{(3)}}
\nc{\Ffive}{F_{(5)}}
\nc{\Fn}{F_{(n)}}
\nc{\Fp}{F_{(p)}}

\nc{\tFzero}{\tF_{(0)}}
\nc{\tFtwo}{\tF_{(2)}}
\nc{\tFfour}{\tF_{(4)}}
\nc{\tFone}{\tF_{(1)}}
\nc{\tFthree}{\tF_{(3)}}
\nc{\tFfive}{\tF_{(5)}}
\nc{\tFn}{\tF_{(n)}}
\nc{\tFp}{\tF_{(p)}}

\nc{\Czero}{C_{(0)}}
\nc{\Ctwo}{C_{(2)}}
\nc{\Cfour}{C_{(4)}}
\nc{\Cone}{C_{(1)}}
\nc{\Cthree}{C_{(3)}}
\nc{\Cfive}{C_{(5)}}
\nc{\Cn}{C_{(n)}}

\nc{\equ}{{\rm eq}}
\def\Im{{\rm Im \hspace{0.5mm} }}

\def\Re{{\rm Re \hspace{0.5mm}}}

\nc{\vol}{{\rm vol}}
\nc{\Ainf}{A_{\infty}}
\nc{\End}{{\rm End}}
\nc{\Ext}{{\rm Ext}}
\nc{\IIB}{{\rm IIB}}
\nc{\Ad}{{\rm Ad}}
\nc{\IIA}{{\rm IIA}}
\nc{\AdS}{{\rm AdS}}
\nc{\CFT}{{\rm CFT}}
\nc{\diag}{{\rm diag}}
\nc{\Log}{{\rm Log}}
\nc{\Mat}{{\rm Mat}}
\nc{\mat}{{\rm mat}}
\nc{\cRic}{{\cR ic}}
\nc{\hcRic}{\widehat{\cRic}}
\nc{\Dir}{{\rm Dir}}

\nc{\Dslash}{\ensuremath \raisebox{0.025cm}{\slash}\hspace{-0.32cm} D}
\nc{\cDslash}{\ensuremath \raisebox{0.025cm}{\slash}\hspace{-0.32cm} \cD}
\nc{\omslash}{\om\!\!\!/}
\nc{\delslash}{\del\!\!\! /}
\nc{\Hslash}{H\!\!\!\! /}
\nc{\Kslash}{K\!\!\!\!/}
\nc{\Yslash}{Y\!\!\!\!/}

\nc{\no}{\!:\!\!}
\nc{\ointdz}{\oint\frac{dz}{2\pi i}}
\nc{\ointdzone}{\oint\frac{dz_1}{2\pi i}}
\nc{\ointdztwo}{\oint\frac{dz_2}{2\pi i}}
\nc{\ointdzb}{\oint\frac{d\zbar}{2\pi i}}
\nc{\ointdzbone}{\oint\frac{d\zbar_1}{2\pi i}}
\nc{\ointdzbtwo}{\oint\frac{d\zbar_2}{2\pi i}}
\nc{\dz}{\frac{dz}{2\pi i}}
\nc{\dzb}{\frac{d\zbar}{2\pi i}}
\nc{\bpm}{\begin{pmatrix}}
\nc{\epm}{\end{pmatrix}}
 \nc{\bitem}{\begin{itemize}}
 \nc{\eitem}{\end{itemize}}
 \nc{\exercise}{\vskip 2mm \noindent {\bf Exercise:}}
 
\begin{document}

\vspace{0.5cm}
\begin{center}
\baselineskip=13pt {\LARGE \bf{Heterotic Hyper-K\"ahler Flux Backgrounds}\\}
 \vskip1.5cm 
Nick Halmagyi$^*$, Dan Isra\"el$^*$, Matthieu Sarkis$^*$ and Eirik Eik Svanes$^{*\dagger}$\\ 
\vskip0.5cm
\textit{
$^\dagger$Sorbonne Universit\'es, Institut Lagrange de Paris, \\
98 bis Bd Arago, 75014, Paris, France,\\
\vskip0.2cm
$^*$Sorbonne Universit\'es, UPMC Paris 06,  \\ 
UMR 7589, LPTHE, 75005, Paris, France \\
\vskip0.2cm
and \\
\vskip0.2cm
$^*$CNRS, UMR 7589, LPTHE, 75005, Paris, France}\\
\vskip0.5cm
halmagyi@lpthe.jussieu.fr \\
israel@lpthe.jussieu.fr \\ 
msarkis@lpthe.jussieu.fr \\ 
esvanes@lpthe.jussieu.fr \\

\end{center}
\begin{abstract}
We study Heterotic supergravity on Hyper-K\"ahler manifolds in the presence of non-trivial warping and three form flux with Abelian bundles in the large charge limit. We find exact, regular solutions for multi-centered Gibbons-Hawking spaces and Atiyah-Hitchin manifolds. In the case of Atiyah-Hitchin, regularity requires that the circle at infinity is of the same order as the instanton number, which is taken to be large. Alternatively there may be a non-trivial density of smeared five branes at the bolt.
\end{abstract}

\section{Introduction}

Heterotic flux backgrounds are interesting models of string backgrounds. There is no in principle impediment from using the R-NS string worldsheet formalism although compact models with minimal or no supersymmetry remain challenging to construct.

In this work we study local Heterotic flux backgrounds on Hyper-K\"ahler four manifolds: in particular the Gibbons-Hawking spaces \cite{Gibbons:1979zt} and the Atiyah-Hitchin manifold \cite{Atiyah:1985dv}. Key to our configurations is that the gauge fields are Abelian and we take the large charge limit such that $\Tr F\w F$ dominates $\Tr R\w R$ in the Bianchi identity. This large-charge limit has been previously studied on the Eguchi-Hanson space \cite{Carlevaro:2008qf} and the conifold \cite{Carlevaro:2009jx, Halmagyi:2016pqu} and this type of limit is familiar from the large-charge supergravity limit crucial to the development of holography \cite{Maldacena:1997re} in type II and M-theory.

Our strategy is to first compute explicit solutions to the Hermitian-Yang-Mills equations on Hyper-K\"ahler spaces and then backreact them on the geometry. This backreaction affects only the conformal mode of the metric but generates a non-trivial three-form flux. According to Aspinwall \cite{Aspinwall:2000fd} we are studying {\it Goldilocks} theories with just the right amount of supersymmetry; perhaps then not surprisingly we solve the supergravity background exactly. For the Atiyah-Hitchin background it is nonetheless somewhat impressive that we can both solve Hermitian-Yang-Mills exactly and integrate analytically the resulting Poisson equation for the backreaction of this instanton. 

The instanton we use on the Atiyah-Hitchin manifold is well known from a classic duality paper by Ashoke Sen \cite{Sen:1994yi}. Perhaps the central result of our work is that the Heterotic backreaction of this instanton can be made regular. This is somewhat non-trivial since the negative mass of Atiyah-Hitchin induces a negative warp factor thus violating the desired signature of space-time. We circumvent this in two ways: first by allowing the asymptotic circle to be large and secondly by including smeared five-brane sources.  

We also study the presence of electric $H$-flux and fundamental strings.The electric flux modifies the BPS equations in a straightforward way and for each solution with magnetic $H$-flux, the electric flux can be added through a harmonic function on the Hyper-K\"ahler manifold. We analyze limits in which we recover AdS$_3$ geometries but these reduce to the known AdS$_3\times S^3\times HK_4$.

Upon completing this work we were made aware that our BPS equations have turned up in five dimensional supergravity. The local equations we study can essentially be found in \cite{Bena:2005va, Bena:2007ju} but with very different global and regularity requirements. This is not surprising since we can dimensionally reduce our solutions on $\RR^5$ to get solutions of ungauged five dimensional supergravity. In addition, these equations also turn up in type II supergravity for $T^2$ fibrations over Hyper-K\"ahler spaces and the type II analogue of the Gibbons-Hawking solutions we find have been analyzed in \cite{Minasian:2009rn}. It is straightforward to convert our solutions on the Atiyah-Hitchin manifold to such type II backgrounds.

\section{Hyper-K\"ahler Heterotic Backgrounds}\label{sec:HKHeterotic}

The primary backgrounds we consider are of the form $\RR^{1,5}\times HK_4$ where $HK_4$ is a warped  Hyper-K\"ahler-four manifold. We consider a non-trivial three-form flux $\cH_{(3)}$, 
dilaton $\Phi$ and Heterotic gauge field $F$. 
The background metric ansatz is:
\bea
{\rm d}s_{10}^2 &=& {\rm d}s_{1,5}^2 + H\, {\rm d}s_{4}^2
\eea
where ${\rm d}s_{4}^2$ is an Hyper-K\"ahler metric on a four-manifold $HK_4$ and $H$ a conformal factor.
The BPS equations are fairly standard:
\begin{subequations}
\label{BPSHK}
\begin{align}
e^{2\phi}\, =&\, H \label{BPSHK1} \\
\cH_{(3)} \, =& \, -*_{4} {\rm d}H \label{BPSHK2} \\
J_a\llcorner F\,=&\, 0\,,\qquad a=1,2,3 \label{BPSHK3}
\end{align}
\end{subequations}
where the $*_{4}$  is the Hodge dual w.r.t. the Hyper-K\"ahler metric on $HK_4$ and $J_a$ 
are the three K\"ahler forms. 

A major difficulty in finding explicit solutions of Heterotic supergravity with non-trivial 
three-form flux is to satisfy the Bianchi identity at the appropriate order in $\alpha'$. Following 
earlier works by some of the authors~\cite{Carlevaro:2008qf,Carlevaro:2009jx,Halmagyi:2016pqu}, our strategy will be 
to work in a large (fivebrane) charge limit, ensuring that the contribution of the $\Tr R\wedge R$ term 
is subdominant and can be consistently neglected. The Heterotic Bianchi identity simplifies to  
\be
{\rm d}\cH_{(3)} = \al' \Tr\, F\w F\label{HetBianchi} 
\ee
implying from the three-form ansatz~(\ref{BPSHK2}) that:
\be
{\rm d}*_{4} {\rm d} H = - \al' \Tr\, F\w F\,.\label{BPSHK4} 
\ee

\subsection{Principal torus bundles and type IIA/IIB solutions} \label{sec:torusbundles}

We consider a more general class of backgrounds, that can be viewed as local models of the principal torus bundles over wrapped K3 surfaces introduced in~\cite{Dasgupta:1999ss} and discussed in many works 
including~\cite{Goldstein:2002pg,Fu:2006vj,Becker:2006et,Becker:2009df}. They generalize the solutions based on Eguchi-Hanson space presented in~\cite{Fu:2008ga}. The general ansatz for such 
principal two-torus bundle $T^2 \hookrightarrow M_6 \stackrel{\pi}{\to} HK_4$ over a Hyper-K\"ahler four-manifold is of the form
\bea
{\rm d}s_{10}^2 &=& {\rm d}s_{1,3}^2 + H\, {\rm d}s_{4}^2 + \frac{U_2}{T_2} |dx + T dy + \pi^\star \alpha|^2\, ,
\eea
where $\alpha$ is a connection one-form on $HK_4$ such that $\vartheta = dx + T dy + \pi^\star \alpha$ is a globally defined one-form on $M_6$ with 
\begin{equation}
\frac{1}{2\pi}  {\rm d} \vartheta = \pi^\star \varpi \ , \quad \varpi = \varpi_1 + T \varpi_2 \ , \quad \varpi_i \in H^2(HK_4,\mathbb{Z})\, ,
\end{equation}
and by supersymmetry 
\begin{equation}
J^a \wedge \varpi = 0 \ , \quad a=1,2,3\, .
\end{equation}
The expression for the three-form becomes then
\begin{equation}
\cH_{(3)} \, = \, -*_{4} {\rm d}H - \frac{\alpha' U_2}{T_2} \text{Re}\, \left( *_{4} {\rm d} \vartheta \wedge \bar\vartheta \right)\, .
\end{equation}

By an appropriate choice of $\varpi \in H^2(HK_4,\mathbb{Z})$ one can find solutions with ${\rm d}\cH_{(3)}  = 0$, which can also be obtained 
as supersymmetric solutions of type IIA or type IIB supergravity with NS-NS fluxes, as was discussed in~\cite{Carlevaro:2008qf} and~\cite{Martelli:2010jx}. 
\section{Gibbons-Hawking: ALE and ALF}

We can solve explicitly~\eq{BPSHK} and~\eq{BPSHK4} for the multicentered Gibbons-Hawking ALE and ALF spaces, that 
we denote collectively by $M_{GH}$. The corresponding Hyper-K\"ahler metrics are given by\footnote{As usual $*_3$ is the Hodge dual on $\RR^3$}: 
\begin{subequations}
\bea
{\rm d}s_{4}^2 &=& V(x)^{-1} ({\rm d}\tau + \om )^2 + V(x) {\rm d}x\cdot {\rm d}x\,, \\
{\rm d}V&=& *_3 {\rm d}\om  \label{dVom}\,, \\
V&=& \eps+ 2m \sum_{i=1}^k \tfrac{1}{|\bx - \bx_i|}\,,
\eea
\end{subequations}
where $\eps=0$ gives the ALE (multi Eguchi-Hanson) series and $\eps=1$  the ALF (multi Taub-NUT) series. The periodicity of $\tau$ is determined by expanding around a pole of $V(\bx)$ to be:
\be
 \tau \sim \tau +  8\pi m \, ,
\ee
and the triplet of K\"ahler forms is given by
\begin{equation}
J_a
= ({\rm d}\tau + \om )\w {\rm d}x^a - V *_3  {\rm d}x^a \quad ,\ a = 1,2,3\, .
\end{equation}

We will consider heterotic supergravity solutions for warped ALE or ALF spaces supported by Abelian gauge bundles. 
To explicitly write the gauge fields we denote
\be
V_i =\frac{2m}{|\bx-\bx_i|}\,,\qquad {\rm d}\om_i = *_3 {\rm d}V_i\, ;
\ee
Then a representative of the topologically non-trivial gauge fields is locally given by
\bea
A_i &=& \om_i - \frac{V_i}{V}({\rm d}\tau + \om)\,.
\eea
We note that 
\be
\sum_{i=1}^k A_i= \eps\frac{{\rm d}\tau+\om}{V}-{\rm d}\tau
\ee
is topologically trivial since $V^{-1}({\rm d}\tau+\om)$ is globally defined. Thus there are $(k-1)$ non-trivial gauge fields, in agreement with the $(k-1)$ non-trivial two-cycles. The corresponding field strengths are
\bea
\label{eq:abel_inst}
F_i={\rm d}A_i 
&=&  V*_3\, {\rm d}\Bslb \frac{V_i}{V}\Bsrb  - {\rm d}\Bslb \frac{V_i}{V}\Bsrb \w ({\rm d}\tau + \om) \,.
\eea
It is straightforward to see that $F_i$ are anti-self dual and solve Hermitian Yang-Mills\footnote{For a one form $\al$ on $\RR^3$ we have $*_4 \al=-(d\tau+\om)\w *_3 \al$. In particular this means that a function which is invariant under the 
$U(1)$ generated by $\del_\tau$ is harmonic on the Gibbons-Hawking space iff it is harmonic on $\RR^3$.}\\
\be
*_4 F_i = -F_i \,,\qquad J_a \w F_j =0\,.
\ee
For the Bianchi identity \eq{BPSHK4} we compute
\bea
F_i \w F_j  &=&-2V  *_3d\Bslb\frac{V_i}{V} \Bsrb \w  d \Bslb \frac{V_j}{V}\Bsrb \w (d\tau + \om)  \\
d *_4 d \Bslb\frac{V_i V_j}{V}\Bsrb&=& -F_i \w F_j 
\eea
so that if we take
\be
F=\frac{1}{4m}\sum_{i=1}^k  dA_i \,\bq_i \cdot \cT\, ,
\ee
where $\cT\in U(1)^{16}$ is in the Cartan subalgebra of $E_8\times E_8$ or $SO(32)$ and $\bq_i$ the corresponding charge vectors, and 
solve~\eq{BPSHK4}, we get the general solution\footnote{we have chosen to work with Hermitian gauge 
fields, normalized as $\Tr\, \cT_\al\cT_{\beta}=2\delta_{\al\beta}$.}:
\be
H= \delta +h(\bx)+ \frac{\al'}{8m^2V} \sum_{i,j=1}^k V_i V_j \bq_i \cdot \bq_j \,. \label{H3GHSol}
\ee 
Here  $\delta=0,1$ is an integration constant (not to be confused with $\eps$ a similar integration constant in the Gibbons-Hawking 
warp factor $V$) and $h(\bx)$ is any harmonic function on $M_{GH}$. Taking $h(\bx)$ to be invariant under $\del_\tau$ we have
\be
h(\bx) =\frac{1}{m} \sum_\al \frac{q_\al}{|\bx-\bx_\al|}
\ee
corresponding to mobile neutral five-brane sources inserted at $\bx_\al$.

In appendix~\ref{app:EguchiHanson} we show how the two center solution is related to the Eguchi-Hanson solution that was 
discussed in particular in~\cite{{Carlevaro:2008qf}}.

\subsection{Five-brane and Magnetic Charges}
At infinity we can compute the five-brane charge using~\eq{BPSHK2} and~\eq{H3GHSol}. 
We have
\bea
\cH_{(3)} 
&=&({\rm d}\tau +\om)\w *_3 {\rm d}H 
\eea
and so\footnote{The volume form of a three-sphere is
\bea
ds_{S^3}^2&=& \frac{1}{4} \Bslb \frac{1}{4m^2}(d\tau+\om)^2 + d\Om_2^2 \Bsrb \non \\
2\pi^2&=&\int \vol(S^3)=  \frac{1}{8}\int \frac{1}{2m}(d\tau+\om)\w\Om_2\,. \non
\eea
}
\begin{subequations}
\bea
{\rm d}H&=& -\frac{\al' }{4mkr^2}\Blp\sum_{i,j=1}^k \bq_i\cdot \bq_j \Brp {\rm d}r -\frac{1}{m}\sum_\al \frac{q_\al}{r^2} +\ldots \\
\cH_{(3)}&=&\Bslb -\frac{\al'}{4mk} \sum_{i,j=1}^k \bq_i\cdot \bq_j -\frac{1}{m}\sum_\al q_\al \Bsrb ({\rm d}\tau+\om)\w \Om_2+\ldots  \eea
\end{subequations}
and the Maxwell five-brane charge is
\bea
\cQ_{M}=\frac{1}{4\pi^2 \al'}\int_{S^3/\ZZ_k} \cH_{(3)} &=&-\frac{2}{k^2} \sum_{i,j=1}^k \bq_i\cdot \bq_j -\sum_\al \frac{8q_\al}{k}\,.
\eea
One can also define a Page charge,  which is quantized, as:
\bea
\cQ_{P}&=& \frac{1}{4\pi^2 \al'}\int_{S^3/\ZZ_k} \Blp \cH_{(3)}-A\w F\Brp \label{Pagedef} \\
&=& -\sum_\al \frac{8q_\al}{k}\in \ZZ  \notag\,.
\eea

\subsubsection*{Magnetic Charges}

We take a basis of two cycles to be $\Delta_{i}$ where the poles of the $\Delta_i$ are at $x_i$ and $x_{i+1}$.
Then the matrix of magnetic charges associated with the Abelian gauge bundle are:
\bea
q_{i,j}=\frac{1}{2\pi} \int_{\Delta_j} F_i&=& \frac{\bq_i\cdot \cT}{8\pi m} \Blp \int {\rm d}\tau\Brp \int _{x_j}^{x_{j+1}}
{\rm d}\Bslb \frac{V_i}{V} \Bsrb  \non \\
&=&  \, \bq_i\cdot \cT \Bslb \delta_{j+1,i} -\delta_{j,i} \Bsrb\,.
\eea

\subsection{Partial blow-down limits and fivebranes}

The function $V(\bx)$ has $k$-poles, now suppose that $k'$ of these poles are co-incident, which correspond 
to a partial blow-down limit of the ALE or ALF space. In the present situation some of the Abelian instantons~\eq{eq:abel_inst} 
become point-like as the corresponding two-cycles shrink and we expect heterotic five-brane to appear. We 
now check that in the region around such a pole we obtain the near horizon of 
five-brane solution of Callan, Harvey and Strominger~\cite{Callan:1991ky, Callan:1991at} 
where the three-sphere is orbifolded by $\ZZ_{k'}$.

For simplicity we set $\bx_j=0$ for $j=1,\ldots k'$, and  in the neighborhood of this pole the functions 
$H$ and $V$ behaves like 
\begin{equation}
H\stackrel{r\to 0^+}{\simeq}\frac{1}{r^2}  \frac{\al'}{2k'm }Q_5 \quad , \qquad 
V \stackrel{r\to 0^+}{\simeq}\frac{4mk'}{r^2}
\end{equation}
hence the solution approaches
\begin{subequations}
\bea
{\rm d}s_{10}^2
&\stackrel{r\to 0^+}{\simeq}&{\rm d}s_{1,5}^2 + 2\al' Q_5 \Bslb \frac{{\rm d}r^2}{r^2}  + \frac{r^2}{4} \left(\sigma_1^2 
+ \sigma_3^2 + (\tfrac{\sigma_3}{2k'm})^2 \right) \Bsrb  \\
\cH_{(3)} &\stackrel{r\to 0^+}{\simeq}&  \frac{\alpha' Q_5}{2} \sigma_1 \wedge \sigma_2 \wedge \tfrac{\sigma_3}{2 k' m}
\eea
\end{subequations}
where the five-brane charge is given by:
\be
Q_5= \sum_{i,j=1}^{k'} \bq_i\cdot \bq_j \,.
 \ee

\subsection{Double Scaling Limit}

For the two-center Eguchi-Hanson solution ($k=2$) there exists an interesting double scaling limit~\cite{Carlevaro:2008qf}, 
defined as:
\be
\label{eq:dsl_limit}
g_s\ra 0\,,\qquad \lambda:=\frac{g_s \sqrt{\al'}}{a} \ {\rm fixed \ and \ finite}\,,
\ee
where $a$ is the distance between the two centers. This limit decouples the asymptotically locally Euclidian region, and $\lambda$ becomes 
the effective coupling constant of the interacting string theory.  

In the spherical coordinates reviewed in appendix~\ref{app:EguchiHanson} the metric of the solution becomes 
\be
{\rm d}s^2={\rm d}s_{1,5} + 
\frac{\al' Q_5}{2}\Bslb \frac{{\rm d}r^2}{r^2(1-\frac{a^4}{r^2})} + \frac{1}{4}\Blp 1- \frac{a^4}{r^2}\Brp  \sig_3^2 +{\rm d}\Om_2^2\Bsrb
\ee
and the corresponding heterotic string theory admits an exactly solvable worldsheet CFT. 
This space has an asymptotic linear dilaton hence admits a holographic description as a 
little string theory~\cite{Aharony:1998ub} but, unlike the CHS background, is given by a smooth solution of heterotic supergravity.

A double scaling limit can be described in principle for arbitrary $k$. Let us define $\mathbf{x}_i = a \mathbf{y}_i$ where the coordinates 
$\mathbf{y}_i$ are dimension-less and $a$ is a common scale factor. The double-scaling limit can be then described exactly as before 
by eq.~(\ref{eq:dsl_limit}); in practice the double scaling limit amounts to setting $\delta\ra 0$ in \eq{H3GHSol}. 

It would be interesting to check if one could derive a worldsheet CFT for the double scaled solutions when $k>2$, 
in particular whenever the centers are arranged following a simple pattern, for instance a homogeneous distribution on a circle.

\section{Atiyah-Hitchin}

The Atiyah-Hitchin space $M_{AH}$ is a four-dimensional smooth manifold with an explicit Hyper-K\"ahler metric which at long distances approximates Taub-NUT with a negative mass parameter. The original work is \cite{Atiyah:1985dv, Atiyah:1985fd} and an interesting simplification was given in \cite{Gibbons:1986df}. Our notation will follow a more recent work \cite{Bena:2007ju} where $M_{AH}$ was used as a potential Euclidean Hyper-K\"ahler  base manifold for five dimensional supergravity solutions. In \cite{Bena:2007ju} regularity required the absence of closed time-like curves and this effectively excluded physical solutions whereas for our computations the non-trivial regularity conditions are essentially just positivity of the warp factor and we will find regular solutions.

The metric is
\bea
ds_{AH}^2&=& \frac{1}{4} a_1^2 a_2^2 a_3^2 d\eta^2 + \frac{1}{4}a_1^2 \sig_1^2+ \frac{1}{4}a_2^2 \sig_2^2+ \frac{1}{4}a_3^2 \sig_3^2
\eea 
with the $SU(2)$ invariant one-forms satifsying $d\sig_i= \frac{1}{2} \eps_{ijk} \sig_j\w \sig_k$ given by
\bea
\sig_1&=& \cos \psi d\tha + \sin \psi \sin \tha d\phi \non \\
\sig_2&=& \sin \psi d\tha -\cos \psi \sin \tha d\phi  \non \\
\sig_3&=& d\psi +\cos \tha d\phi \non
\eea
and the $a_i$ are subject to the following system of ODE's:
\bea
\frac{\dot{a_1}}{a_1}&=& \frac{1}{2} \Bslb(a_2-a_3)^2 -a_1^2\Bsrb
\eea
and cyclic permutations (dot is the derivative with respect to $\eta$).
One defines new functions quadratic in the $a_i$:
\be
w_1 =a_2 a_3\,,\quad w_2=a_1a_3\,,\quad w_3 =a_1a_2
\ee
and then the system of ODE's is then
\bea
(w_1+w_2)'&=& -2\frac{w_1w_2}{u^2} \\
(w_2+w_3)'&=& -2\frac{w_2w_3}{u^2} \\
(w_3+w_1)'&=& -2\frac{w_3w_1}{u^2}
\eea
(prime is derivative with respect to $\theta$) with solution
\bea
w_1&=& -uu' -\frac{1}{2}u^2\csc\tha \\
w_2&=& -uu' + \frac{1}{2}u^2\cot \tha \\
w_3&=& -uu' + \frac{1}{2}u^2 \csc\tha \,.
\eea
where
\be
u=\frac{1}{\pi} \sqrt{\sin\tha} K(\sin^2 \frac{\tha}{2})
\ee
and $\eta$ is given in terms of $\tha$ through
\be
 u^2 d\eta= d\tha\,,\qquad \eta= -\int^\pi_\tha \frac{d\tha}{u^2}\,.
\ee

For our gauge field ansatz we need an anti-self dual two form on $M_{AH}$, this is then guaranteed to solve Hermitian Yang-Mills without the need to construct the explicit Hyper-K\"ahler structure\footnote{One could in principle write down the Hyper-K\"ahler structure using the results of \cite{Olivier1991} or by computing the Killing spinors.}. In a classic paper on dualities \cite{Sen:1994yi}, Sen gave an integral expression for exactly such an anti-self dual, harmonic two-form on $M_{AH}$ but the appearance of this two-form dates back to the works \cite{Manton:1987pa, Gibbons:1987sp, Manton:1993aa}. Interestingly, from the work \cite{Bena:2007ju} we have the closed-form expression of this two-form
\bea
\Om &=& h \blp a_1^2 dr\w \sig_1 - \sig_2\w \sig_3 \brp \,,\label{Omdef}\\
h&=&\frac{u^2}{w_1\sin \frac{\tha}{2}}\,.
\eea
In \cite{Bena:2007ju} they consider self-dual forms but with a small modification of the frames this is made anti-self dual. More precisely our choice of frames is
\be
e_0=\frac{a_1a_2a_3}{2} d\eta\,,\qquad e_i =\frac{a_i}{2} \sig_i\,,
\ee
whereas in \cite{Bena:2007ju} an additional minus sign in $e_0$ was used. So we have locally
\be
\Om= -d(h\, \sig_1)\,.
\ee
In fact one can construct a triplet of anti-self dual forms $\Om_-$ and a triplet of self-dual forms $\Om_+$ in a similar manner:
\be
\Om_{i-}=-d(h_i \sig_i)\,,\qquad \Om_{i+}= d(h_i^{-1}\sig_i)
\ee
with
\be
h_1=\frac{u^2}{w_1\sin \frac{\tha}{2}}\,,\quad h_2=\frac{u^2}{w_2}\,,\quad h_3=\frac{u^2}{w_3\cos\frac{\tha}{2}}\,,
\ee
however only $\Om_{1-}=\Om$ is normalizable. Given that there is a single non-trivial two-cycle in $M_{AH}$ one might be pleased to know that this normalizable form is dual to this two-cycle but there was no guarantee that the dual two-form would have an $SU(2)$ invariant representative.

\subsection{Bianchi Identity}

We take our gauge field to be
\be
F=\Om\, \bq\cdot \cH\, \label{Fdef}
\ee
where $\cH\in U(1)^{16}$ is in the Cartan subalgebra of $E_8\times E_8$ or $SO(32)$ and $\bq$ the corresponding charge vector. The three-form flux is
\bea
\cH_{(3)} 
&=& - \frac{H'}{4} \sig^{1}\w \sig^2 \w \sig^3 \non \\
\eea
and the Bianchi identity is
\be
d*_4 d H = -2q^2 \Om\w \Om\,, \label{AHBianchi}
\ee
where $q^2=\bq\cdot \bq$.

Quite remarkably, one can integrate this Poisson equation explicitly
\be
H= h_0 + h_1 \eta + \frac{2 q^2}{w_1}
\ee
where  $\{h_0,h_1\}$ are constant coefficients of the s-wave harmonic functions on $M_{AH}$. The last term is manifestly negative definite for the whole region $0\leq \tha \leq \pi$ but we will see that one can compensate for this by a choice of harmonic function and obtain a positive definite warpfactor.

\subsection{Regularity}
The regularity of $M_{AH}$ has been previously studied in detail, we repeat it here to help determine regularity of our warp factor.

In the region $\tha\sim \pi$, we define a radial co-ordinate $r=-\log \cos \frac{\tha}{2}$
and using
\bea
K&=& r +\log(4) +\ldots \\
u&=& \frac{\sqrt{2}}{\pi} r e^{-r/2} +\ldots \\
w_1&=&-\frac{r}{\pi^2 }
\eea
we find that the metric is
\be
ds_{AH}^2 = dr^2 + r^2 (\sig_1^2+\sig_2^2)+ \sig_3^2 + \ldots  \label{metUV}
\ee
and 
\bea
\frac{1}{w_1} &=& -\frac{\pi^2}{r} + \cO(r^{-2}) \\
\eta &=& - \frac{\pi^2}{r}+ \frac{\pi \log (4)}{r^2}+  \cO(r^{-3})
\eea
so that the asymptotic expansion of the warp factor is
\be
H= h_0 -\frac{h_1+2q^2}{r} +\ldots \label{HUV}
\ee

In the region $\tha\sim 0$, we define a new radial variable $\rho =\frac{\tha^2}{64}$ and the metric is
\be
ds_{AH}^2 = d\rho^2 + 4\rho^2 \sig_1^2+\frac{1}{16} ( \sig_2^2+ \sig_3^2) + \ldots  \label{metIR}
\ee
with
\bea
\frac{1}{w_1}  &=&  -4 + 32 \rho^2 + \ldots \\
\eta&=&  \log \rho^2  +\ldots
\eea
so that the IR expansion of the warp factor is
\be
H= (h_0-8q^2) +64q^2 \rho^2+\ldots
\ee
From these expansions we see that with
\be
h_0 > 8q^2 \,,\qquad h_1=0
\ee
we have a positive warp factor which is regular everywhere. We define a rescaled radial coordinate near $\tha\sim\pi$ to be $\hr = h_0^{1/2}r$  and $\hrho=h_0^{1/2} \rho$ near $\tha=0$ so that
\bea
\tha\sim\pi:\qquad ds_{10}
&=& ds_{1,5}^2 +  d\hr^2 +\hr^2 (\sig_1^2+\sig_2^2)+ h_0\sig_3^2  +\ldots \\
\tha\sim0:\qquad ds_{10}^2&=& ds_{1,5}^2 +  d\hrho^2 + 4\hrho^2 \sig_1^2+\frac{h_0}{16} ( \sig_2^2+ \sig_3^2)   +\ldots
\eea
and see that the cost of a positive warp factor is that both the circle at infinity and the two-sphere at the bolt are large. 

It is also important that the $\Tr R_-\w R_-$ term in the Bianchi identity remains small compared to $\Tr F\w F$. From explicit computations we find that the only possible divergences in $\Tr R_-\w R_-$ appear through the warp factor as\footnote{An explicit computation using the Chern connection can be found in \cite{Fu:2006vj} and agrees with our conclusion here.} $\{H'/H\,, H''/H\}$
which by tuning $h_0$ can be made sufficiently small with respect to $\Tr F\w F$. This confirms that our large charge approximation remains valid and these warped Atiyah-Hitchin solutions are good Heterotic backgrounds at leading order.

Alternatively we can obtain a positive warp factor through\footnote{The value of $h_0$ could be chosen to be another non-zero number $\cO(q^0)$.} 
\be
h_0=1,\qquad h_1<-2q^2\,. \label{smearingh01}
\ee
This corresponds to smearing neutral five-branes on the $S^2$ at $\tha=0$. Note that due to this smearing, at the IR ($\tha=0$) the harmonic function parameterized by $h_1$ scales like a source in $\RR^2$. In the UV ($\tha=\pi$), due to the finite circle, the harmonic function scales like $\frac{1}{r}$ which is that of a source in $\RR^3$ not $\RR^4$. The solution is of course singular for the usual reason that smeared branes are singular but this is of a good type and is resolved in string theory.

\subsection{Five-brane Charge}
Computing the five-brane charge requires understanding some global features of $M_{AH}$. From \eq{metUV} and \eq{metIR} we see that there are two inequivalent, emergent $U(1)$ symmetries in the UV and IR, which are broken in the bulk. From \cite{Gibbons:1986df} we know that a regular manifold requires the periodicities to be
\be
0\leq \psi\leq 2\pi,\quad 0\leq \tha \leq \pi\,,\quad 0\leq  \phi\leq 2\pi 
\ee
as well as that the free $\ZZ_2$ symmetry 
\be
I_1: \quad \tha\ra \pi -\tha \,,\quad \phi\ra \pi + \phi\,,\quad \psi \ra -\psi \label{I1}
\ee
is enforced. The horizontal space in the UV is thus $\RR\PP^3/I_1$.

Using \eq{HUV} we have
\be
\cH_{(3)}=\Bslb (h_1 + 2q^2)+\ldots \Bsrb\w \sigma^1\wedge\sigma^2\wedge\sigma^3
\ee
compute the Maxwell five-brane charge to be
\be
\cQ_{M}=\frac{1}{4\pi \al'} \int_{\RR\PP^3/I_1} \cH_{(3)} = h_1+2q^2 \,.
\ee
This is not required to be quantized. The Page charge is defined as in \eq{Pagedef} and we find
\be
\cQ_P=h_1
\ee
which must be integral.

\subsection{Gauge Field Charge}

The gauge field charge is computed using \eq{Omdef} and \eq{Fdef} and the IR expansion 
\be
h=-2 + \ldots 
\ee
Under the symmetry \eq{I1}, the bolt remains a two sphere\footnote{As explained in \cite{Gibbons:1986df} there is an additional, optional $\ZZ_2$ symmetry usually denoted $I_3$ which would convert the bolt into an $\RR\PP^3$} whose volume is $4\pi$. We find
\bea
\frac{1}{2\pi}\int_{S^2} F 
&=& 2\bq\cdot\cH \frac{1}{2\pi} \int_{S^2} \sig_2\w \sig_3 \non \\
&=& 4\bq\cdot\cH\in \ZZ\,.
\eea

\section{Fundamental String Sources and AdS$_3$ Solutions} \label{sec:fstrings}

Heterotic backgrounds with an $\RR^{1,1}$ factor allow for the inclusion of F1-strings along $\RR^{1,1}$ in addition to  the magnetic five-branes. The electric source of three form flux induces a non-trivial warp factor and allows for AdS$_{3}$ solutions. To include these fundamental strings, we first consider internal eight-manifolds $X_8$ and then specialize the internal manifold to be a product of Hyper-K\"ahler manifolds.

The metric and three form are
\bea
ds_{10}^2 &=& e^{2A} ds_{1,1}^2   + ds_8^2  \\
H^{(3)}&=& \vol_2 \w h^{(1)} + h^{(3)}
\eea
where $\vol_2= e^{2A} dx^0 \w dx^1$. Then we find that the BPS equations are a slight embellishment of those found in  \cite{Gauntlett:2003cy}:
\bea
 \Psi\llcorner d  \Psi &=& d(\phi- A) \label{WarpBPS1} \\
H_{(3)}&=& h_{(1)} \w \vol_2 + h_{(3)}\label{WarpBPS2}   \\
h_{(1)}&=& -2dA  \label{WarpBPS3}  \\
h_{(3)}&=&  *_8 e^{2(\phi-A)} d \Blp  e^{2(A- \phi)}  \Psi\Brp \label{WarpBPS4} 
\eea
where $\Psi$ is the $Spin(8)$ structure on $M_8$
\footnote{We note that with canonical holomorphic frames $E_i=e_{2i-1}+i e_{2i}$ such that $ds_8^2 = E_i\otimes \Ebar_i$ the $SU(4)$ structure is $ J= \frac{1}{2i} E_i \w \Ebar_i\,,\ \ \Om=E_1\w E_2 \w E_3 \w E_4\ $
and 
\be
\Psi= \frac{1}{2} \Blp J\w J + \Om + \Ombar \Brp\,.
\ee
}
:
\bea
\Psi&=&e^{1234} + e^{1256} + e^{1278} + e^{3456} + e^{3478} + e^{5678} \non \\
&& + e^{1357} -e^{1368} -e^{1458} -e^{1467} -e^{2358} -e^{2367}-e^{2457} + e^{2468}\,.
\eea
We must supplement the BPS equations with the Bianchi identity \eq{HetBianchi} and then due to the non-trivial warp-factor $A$, one must also impose the three form flux equation of motion:
\bea
 0= d\Blp e^{-2\phi} *_{10} H_{(3)}\Brp \qquad \Ra \qquad
 \begin{cases} 
 0 = d\blp e^{-2\phi} *_8 h_{(1)}\brp\\ 
 0= d\blp e^{2(A-\phi)} *_8 h_{(3)} \brp
 \end{cases}\,.
\eea
\subsection{Product of Hyper-K\"ahler Manifolds}

Our solutions with string and five-brane charges have a natural splitting of the internal eight manifold into a product of Hyper-K\"ahler manifolds\footnote{One might consider an additional warp factor in front of $ds_{M_1}^2$ however from \cite{Strominger:1986uh,Hull:1986kz} we know that this must be constant.}
\be
ds_{10}^2= e^{2A} ds_{\RR^{1,1}}^2 +ds^2_{M_1} +e^{2B} ds^2_{M_2} \label{hyperprod}
\ee
where $M_i$ are both HyperKahler four manifolds, whose triplet of K\"ahler forms we denote
\be
 \{J_i,\Re\Om_i,\Im\Om_i\}\,.
 \ee 
 The functions $A,B$ depend only on the co-ordinates $y_i$ of $M_2$. The $Spin(8)$ structure is given by
\be
\Psi= \frac{1}{2} \Blp J_1\w J_1 +2 e^{2B} J_1 \w J_2 + e^{4B} J_2 \w J_2 + e^{2B} (\Om_1\w \Om_2 + \Ombar_1 \w \Ombar_2 )   \Brp\,.
\ee

We find the BPS conditions, Bianchi identity and equations of motion give\footnote{Note that is $E=e^B\tE$ is an 8d frame $*_8 E \w J_1\w J_1=2*_8 E \w \vol_{M_1}= 2e^{3B} *_{M_2}\tE = 2e^{2B} *_{M_2}E$}
\bea
\phi&=&A+B  \label{HyperBPS1}\\
h_{(3)}
&=&  -*_{M_2}  d \, e^{2B} \\
 d*_{M_2}  d\, e^{2B} &=&-\frac{1}{2} \alpha' \Tr F\w F \label{HyperBPS3}\\
0 
&=& d *_{M_2} d\, e^{-2A}  \\
0&=& J_2 \llcorner F\label{HyperBPS5}
\eea
so we see that the only additional pieces of data from the equations in section \ref{sec:HKHeterotic} is that $e^{2A}$ is harmonic on $M_2$ and the dilaton receives a shift proportional to $A$. For the Atiyah-Hitchin manifold we can smear F1-strings on the $S^2$ bolt in much the same way as we have described for smearing 5-branes on the bolt around \eq{smearingh01}, that is by 
\be
e^{2A}\sim \eta\,.
\ee
We will now be somewhat more explicit for the Gibbons-Hawking spaces.

\subsection{AdS$_3$ from Gibbons-Hawking}
When $M_2$ is a Gibbons-Hawking space, the $U(1)$ invariant harmonic functions are
\be
e^{-2A}=1+\sum_{r} \frac{\hq_r}{|\bx-\bx_r|}
\ee 
corresponding to strings placed along $\RR^{1,1}$ and at fixed points of $\del_\tau$ on $M_2$. 

If in addition we choose to place these strings at poles of $V$ we recover AdS$_3$ geometries near such a pole. We  put $k'$ poles of $V$ as well as the strings at $\bx_r=\bx_i=0$ then in the vicinity of $x_i$ we  have
\bea
e^{2A}&=&\frac{r}{\hq_0},\ldots \quad e^{2B}=\frac{1}{r}  \frac{\al'}{4m }Q_5 +\ldots\,,\qquad V =\frac{2mk'}{r} +\ldots \\
\eea
so that
\bea
ds_{10}^2
&=&\frac{r}{\hq_0}ds_{1,1}^2++ ds_{M_1}^2  + 2\al' k'^2 Q_5 \Bslb \frac{1}{4} \frac{dr^2}{r^2}  +  ds_{S^3/\ZZ_k}^2 \Bsrb  \\
&=& 2\al'k'^2 Q_5\Bslb ds_{AdS_3}^2+ ds_{S^3/\ZZ_k}^2 \Bsrb + ds_{M_1}^2 \\
e^{2\phi} &=& \frac{\al' Q_5}{4m\hq_0}
\eea
where $r=\rho^2$\,. The F1-charge is given as usual by 
\be
Q_1=\frac{4m \hq_0 \vol(M_1)}{\al'^3}\,.
\ee
The gauge field vanishes in this limit and the background is sourced by three-form flux. 

\section{Conclusions}

The key aspect of our solutions with Abelian gauge bundles is that we have taken a large charge limit and consistently suppressed the 
$\Tr\, R\w R$ term in the Bianchi identity, which is subdominant at leading order in the expansion in $\frac{1}{q^2}$.  
We have shown how this large charge limit  can lead to exact supersymmetric flux backgrounds. and it is particularly interesting the the Atiyah-Hitchin manifold can provide a regular background. This configuration requires some ingenuity to counteract the negative mass and result in a background of the correct signature. This Atiyah-Hitchin based solution is distinctly different from those based on Gibbons-Hawking; while the latter can be viewed as marginal deformations of the orbifold of the CHS solutions the finite two-cycle in the Atiyah-Hitchin manifold cannot be blown down. As such we do not have a worldsheet theory from which we can imagine obtaining this as the background geometry.

In these backgrounds, the gauge fields are completely solved for by using the Hermitian Yang-Mills equations which then provide a source for the three-form flux. It is conceivable that non-Abelian bundles could be constructed such that $\Tr\, F\w F$ dominates $\Tr\, R\w R$ everywhere\footnote{Interesting five dimensional solutions with non-Abelian gauge fields have appeared recently \cite{Meessen:2015enl} and the lift to the Heterotic string has been discussed \cite{Cano:2017qrq}. However it is not clear to us how these solutions will solve the exact Bianchi identity}. Since the Kronheimer-Nakajima construction~\cite{Kronheimer1990} gives a solution of all instantons on ALE Gibbons-Hawking spaces, one could possibly even construct such instantons, however most instantons will provide a source the Bianchi identity whose solution is a general function of four variables and thus unsolvable. A particularly neat class of instantons is based on the 't Hooft ansatz~\cite{BoutalebJoutei:1979iz}:
\bea
A_0&=&\frac{1}{2}\vG \cdot \vecsig\,,\qquad \vA=\frac{1}{2}\Bslb \vom (\vG \cdot \vecsig) - V(\vG\times \vecsig)\Bsrb \non \\
\vG&=& - V^{-1} \vec{\nabla}\log f
\eea
with $f$ harmonic on $\RR^3$. For finite action, the centers of $f$ are constrained to lie at the poles of $V$. These instantons can have large $\Tr\, F\w F$ in the limit of large number of poles of $f$ but $\Tr\, R\w R$ will not be suppressed.

There are numerous directions for progress on the worldsheet description of these backgrounds. The elliptic genus for type II on ALE spaces has been computed recently \cite{Harvey:2014nha} based on general developments in this field \cite{Benini:2013xpa} and we expect to be able to provide a similar solution for these Heterotic models or the type II flux backgrounds of section \ref{sec:torusbundles}. It would also be interesting to provide an exactly solvable worldsheet model of the near-horizon region of the multi-center Gibbons-Hawking backgrounds, generalizing the gauged WZW model of the two-centered solution.

\vskip 1cm

\noindent {\bf Acknowledgements} N.H. would like to thank Nikolay Bobev and Davide Cassani for discussions. E.S. would like to thank Bobby Acharya for insightful conversations. This work was conducted within the ILP LABEX (ANR-10-LABX-63) supported by French state funds managed by the 
ANR within the {\it Investissements d'Avenir} program (ANR-11-IDEX-0004-02), by the project QHNS in the program ANR Blanc 
SIMI5 of Agence National de la Recherche and the CEFIPRA grant 5204-4.

\begin{appendix}
\section{Eguchi-Hanson}\label{app:EguchiHanson}

When $k=2$ and $\eps=0$, the explicit co-ordinate transformation is known \cite{Prasad:1979kg} from the Gibbons-Hawking space to the Eguchi-Hanson space \cite{Eguchi:1978xp}. In Cartesian co-ordinates the two center ALE Gibbons-Hawking space has
\bea
\om &=& \Bslb \frac{z-a^2/8}{\sqrt{x^2 + y^2 + (z-a^2/8)^2}} + \frac{z+a^2/8}{\sqrt{x^2 + y^2 + (z+a^2/8)^2}} \Bsrb d\blp \tan^{-1} \frac{y}{x}\brp  \\
V&=& \frac{1}{\sqrt{x^2 + y^2 + (z-a^2/8)^2}}+\frac{1}{\sqrt{x^2 + y^2 + (z+a^2/8)^2}}\,.
\eea
Following \cite{Prasad:1979kg} we have $(a\leq r)$:
\bea
x&=& \frac{a^2}{8} \sqrt{\frac{r^4}{a^4}-1} \sin \tha \cos \psi \non \\
y&=& \frac{a^2}{8} \sqrt{\frac{r^4}{a^4}-1}\sin \tha \sin \psi \non \\
z&=&\frac{1}{8}r^2 \cos \tha \non
\eea
so that
\bea
V
&=& \frac{16}{a^2}\frac{ \frac{ r^2}{a^2 }}{\frac{r^4}{a^4}- \cos^2 \tha }\,, \non \\
\om &=& \frac{2 \cos \tha (\frac{r^2 }{a^2}-1)}{\frac{r^2 }{a^2}-\cos \tha} d\psi\,.
\eea

As an example, we write explicitly the solution for Heterotic five-branes on Eguchi-Hanson with additional F1-strings\footnote{One can take $M_1$ to be $T^4$ or $K3$ with the Ricci-flat metric}. 

\bea
ds_{M_2}^2&=& \frac{dr^2}{f^2} + \frac{r^2}{4} \bslb \sig_1^2+\sig_2^2+ f^2\sig_3^2\bsrb \\
f^2&=& 1-\frac{a^4}{r^4} \\
h_{(3)}&=& 2 f^2 r^3(e^{2B})' \sig_1\w \sig_2\w \sig_3 \non \\
F&= & d\Blp \frac{a^2}{r^2} \eta \Brp \\
e^{2B} &=& 1+\frac{8\al' q^2}{r^2 }+\frac{Q_5}{8 a^2} \log\Bslb \frac{r^2/a^2 - 1}{r^2/a^2 + 1}\Bsrb   \\
e^{-2A} &=&1+\frac{Q_1}{a^2} \log\Bslb \frac{r^2/a^2 - 1}{r^2/a^2 + 1}\Bsrb \\
e^{2\Phi} &=& e^{2(A+B)}
\eea

In addition to the Heterotic five-branes which resolve the singularity, there are $Q_1$ mobile F1-strings and $Q_5$ NS5-branes smeared on the blown-up $S^2$. Due the the smearing of the strings, the near horizon limit has a log-singularity at $r\sim a$ in the warp factor $e^{2A}$ and thus there is no enhancement to AdS$_3$.
In the blow-down limit $a\ra0$ where the Eguchi-Hanson space becomes $\CC^2/\ZZ_2$, the gauge field vanishes and we get the $\ZZ_2$ orbifold of the usual F1-NS5-solution, the near-horizon limit is AdS$_3\times S^3/\ZZ_2\times M_1$.

\end{appendix}

\providecommand{\href}[2]{#2}\begingroup\raggedright\endgroup

\end{document}